\documentclass[12pt]{iopart}
\usepackage{iopams}
\begin{document}

\title{Relativistically invariant extension of the de Broglie-Bohm theory of
quantum mechanics}
\author{Chris Dewdney and George Horton}

\begin{abstract}
We show that quantum mechanics can be given a Lorentz-invariant
realistic interpretation by applying our recently proposed
relativistic extension of the de Broglie-Bohm theory to deduce
non-locally correlated, Lorentz-invariant individual particle
motions for the EPR experiment and the double-interferometer
experiment proposed by Horne, Shimony and Zeilinger.
\end{abstract}

 \maketitle

\address{Division of Physics, University of Portsmouth. Portsmouth PO1 2DT.
England}

\section{Introduction}

In the de Broglie-Bohm approach to quantum theory configuration space is the
fundamental arena. In non-relativistic quantum theory, given the wave
function of an $n$-particle system $\Psi \left( \overrightarrow{r}^{(1)},%
\overrightarrow{r}^{(2)}....\overrightarrow{r}^{(n)};t\right),$
where the bracketed superscript labels the particle, and some
``initial'' point in the configuration space-time, the entire
motion of the system in space and time is determined. Each initial
point in the configuration space-time determines one possible
motion of the system, but, of course, we cannot control which
trajectory actually occurs in a given case. With entangled wave
functions the velocity of a given particle at a particular point
will depend on the positions of all the other particles in the
system. Naturally, in a non-relativistic theory one considers the
positions of the particles at the same time. However, in
relativistic, multi-time quantum theory there is no universal time
coordinate and in a given inertial reference frame, $\Sigma ,$
the argument of the wave function may be written $\left( \overrightarrow{r}%
^{(1)},t^{(1)}....\overrightarrow{r}^{(n)},t^{(n)}\right) ,$ using
an appropriate coordinate system$.$ Just as in the
non-relativistic theory, to obtain a definite trajectory for the
relativistic system an ``initial'' point in the configuration
space-time must be specified. Also like the nonrelativistic case,
we cannot control which initial point is the actual point in a
given run of an experiment. However, as we have discussed in
detail elsewhere (\cite{horton2002}, \cite{horton2000}), and
unlike the non-relativistic case, the specification of this point
alone, although necessary, is not sufficient to determine a unique
configuration space-time trajectory of the system. The problem is
that with entangled wave functions the individual particle
velocities depend on all the arguments of the wave function; so,
in addition to their initial values, one must specify a rule which
coordinates the individual arguments in the wave function beyond
their initial coordinated values, in order to integrate the
equations of motion. It is important to be clear that the
coordination that we are discussing is not the coordination of
points on given, pre-existing individual particle world lines. The
particle world lines are not specifiable at all unless one has
decided how the particle coordinates are to be coordinated beyond
the initial chosen point. The choice of the initial point is
arbitrary, just as as in the non-relativistic theory but, just as
in the non-relativistic theory, each choice can be associated with
a certain probability density. In general there will be some set
of points which corresponds with the particular experimental set
up being considered. Different choices of
initial point, $\left( \overrightarrow{r}^{(1)},t^{(1)}....\overrightarrow{r}%
^{(n)},t^{(n)}\right) ,$ will lead to different trajectories. Furthermore,
even given the same initial point, different rules of coordination of the
particle coordinates will lead to different configuration space-time
trajectories.

Several suggestions have been made concerning the coordination of the
particle coordinates. For example, Bohm and Hiley \cite{Bohm} have proposed
that there exists a preferred frame of reference in which one takes equal
time steps for each particle's time coordinate in integrating the equations
of motion$.$ In Bohm's approach relativistic invariance appears only as a
property of the statistical results of measurement whilst the individual
processes themselves are not invariant. Suarez \cite{Suarez} has proposed a
``multisimultaneity'' theory in which the measuring devices determine the
coordination of the particles along their simultaneity hyperplanes, but this
theory has experimental consequences different to quantum mechanics and has
in fact been refuted \cite{Gisin}. Other approaches\cite{durr}, \cite{ghose}%
, \cite{holland}, although based on four-velocities, have used an arbitrary
foliation of space-time to provide a basis for integrating the equations of
motion; but then the particle trajectories for given initial conditions,
calculated using different foliations are not lorentz transforms of each
other. To the extent that the positions and momenta depend on the choice of
foliation a theory cannot be considered a theory of ``beables''.

Our approach \cite{horton2002} is based on the existence of time-like four
velocities, for both bosons and fermions, and uses a relativistically
invariant rule, utilizing the invariant light-cone structure, to produce the
system trajectory. In our approach the coordination of the coordinates is
achieved by advancing the arguments in the wave function so that, for all $n$
\[
\left( \Delta t^{(n)}\right) ^{2}-\left( \Delta x^{(n)}\right) ^{2}-\left(
\Delta y^{(n)}\right) ^{2}-\left( \Delta z^{(n)}\right) ^{2}=\Delta \tau ^{2}
\]
Our approach yields a unique and relativistically invariant trajectory in
the configuration space-time of the system. The configuration space-time
trajectory determines both the individual world lines of the particles and
the coordination of the points on the individual particle world lines. The
motion is irreducibly defined in the configuration space-time spanned by $%
\left( \overrightarrow{r}^{(1)},t^{(1)}....\overrightarrow{r}%
^{(n)},t^{(n)}\right) $ and a line in this space determines the sets of
values of the coordinates that are coordinated (taken together in the
calculation of the trajectory). The system's motion is determined once an
``initial'' point in the configuration space-time is specified and our use
of the word ``initial'' merely indicates a starting point for the
calculation. A particular experiment will be consistent with some set of
initial points, distributed with the appropriate density in the
configuration space-time, which determine the possible motions of the
system. Initial points, not within the appropriate set in configuration
space-time may be chosen but, in general, these will correspond to a
different experiment.

Just as in non-relativistic de Broglie-Bohm theory there is no wave-packet
collapse in our relativistic extension. Measurements play no fundamental
role, they are simply interactions between systems during which a
correlation is introduced between their variables such that by observing one
variable one can infer the value of the other. The inclusion of the
additional system requires the enlargement of the configuration space-time
in which the measurement must be described as a dynamical process relating
system and measuring device coordinates.

Since the system trajectory in configuration space-time is calculated in a
relativistically invariant way it cannot matter which frame of reference we
choose merely to describe the system's motion; changing frames of reference
simply amounts to a passive re-assignment of coordinates. The outcome of any
specific experiment, for a given ``initial'' point in the configuration
space-time, will be the same in all frames. If, in a given experiment, parts
of the apparatus are in relative motion (such as the beam splitters in the
experiment of Stefanov et al\cite{Gisin}) it cannot matter whether we choose
the rest frame of one moving part or that of another, the system's motion,
for a given initial point in configuration space-time, will be the same.

In the following section we consider how our approach can be applied to the
case of spin entanglement in an EPR experiment. We then go on to illustrate
our approach in the context of the two-particle experiment proposed by Horne
et al. \cite{HSZ}.

\section{EPR experiment}

Consider the Einstein-Podolsky-Rosen (EPR) experiment, in the form first
proposed by David Bohm, in which two (space-like separated) spin-one-half
particles are entangled in a total spin zero state. The spinor wave
function, in an inertial frame $\Sigma $ and using an appropriate coordinate
system, may be written
\[
f_{1}\left( \overrightarrow{r}^{\left( 1\right) },t^{\left( 1\right)
}\right) f_{2}\left( \overrightarrow{r}^{\left( 2\right) },t^{\left(
2\right) }\right) \left[ \left(
\begin{array}{c}
1 \\
0
\end{array}
\right) ^{_{\left( 1\right) }}\left(
\begin{array}{c}
0 \\
1
\end{array}
\right) ^{\left( _{2}\right) }-\left(
\begin{array}{c}
0 \\
1
\end{array}
\right) ^{_{\left( 1\right) }}\left(
\begin{array}{c}
1 \\
0
\end{array}
\right) ^{\left( _{2}\right) }\right]
\]
where $f_{1}$and $f_{2}$ are localized wave packets moving in opposite
directions along the $x$ axis and the bracketed superscript labels the
particles. Figure 1 shows the particle paths in a space-time diagram of the
experiment in the $\Sigma $ frame. At time $t_{i},$ in $\Sigma ,$ the $z$
components of the spins of the particles are measured by Stern-Gerlach
apparatuses yielding the wave function, for both $t^{\left( 2\right) }$ and $%
t^{\left( 1\right) }>t_{i}.$%
\[
\left[ f_{1}^{+}\left( \overrightarrow{r}^{\left( 1\right) },t^{\left(
1\right) }\right) f_{2}^{-}\left( \overrightarrow{r}^{\left( 2\right)
},t^{\left( 2\right) }\right) \left(
\begin{array}{c}
1 \\
0
\end{array}
\right) ^{_{\left( 1\right) }}\left(
\begin{array}{c}
0 \\
1
\end{array}
\right) ^{\left( _{2}\right) }-f_{1}^{-}\left( \overrightarrow{r}^{\left(
1\right) },t^{\left( 1\right) }\right) f_{2}^{+}\left( \overrightarrow{r}%
^{\left( 2\right) },t^{\left( 2\right) }\right) \left(
\begin{array}{c}
0 \\
1
\end{array}
\right) ^{_{\left( 1\right) }}\left(
\begin{array}{c}
1 \\
0
\end{array}
\right) ^{\left( _{2}\right) }\right]
\]
where $f^{+}$ and $f^{-}$ are wave packets moving additionally with opposite
momenta along the direction of the Stern Gerlach fields. In order to
calculate trajectories and spin-vector orientations for the particles one
needs a definition of the velocity and spin as beables and, as we have
already emphasized, a rule for coordinating the system's coordinates. A
hidden-variable theory for the Dirac equation was provided by Bohm in 1953
\cite{Bohmdirac}. In order to begin the calculation of the trajectory of the
system in its eight-dimensional configuration space-time one needs to choose
an initial position $\left( \overrightarrow{r_{0}}^{\left( 1\right)
},t_{0}^{\left( 1\right) },\overrightarrow{r}_{0}^{\left( 2\right)
},t_{0}^{\left( 2\right) }\right) $. Then, given this initial coordination
of the particle coordinates, our Lorentz-invariant rule can be applied to
calculate the configuration space-time trajectory and associated
individual-particle spin components. The result of the calculation can be
displayed in a ``look-up'' table as follows:
\[
\begin{array}{cccccc}
\overrightarrow{r}_{0}^{\left( 1\right) } & t_{0}^{\left( 1\right) } &
\overrightarrow{s}_{0}^{\left( 1\right) } & \overrightarrow{r}_{0}^{\left(
2\right) } & t_{0}^{\left( 2\right) } & \overrightarrow{s}_{0}^{\left(
2\right) } \\
\overrightarrow{r}_{1}^{\left( 1\right) } & t_{1}^{\left( 1\right) } &
\overrightarrow{s}_{1}^{\left( 1\right) } & \overrightarrow{r}_{1}^{\left(
2\right) } & t_{1}^{\left( 2\right) } & \overrightarrow{s}_{1}^{\left(
2\right) } \\
\overrightarrow{r}_{2}^{\left( 1\right) } & t_{2}^{\left( 1\right) } &
\overrightarrow{s}_{2}^{\left( 1\right) } & \overrightarrow{r}_{2}^{\left(
2\right) } & t_{2}^{\left( 2\right) } & \overrightarrow{s}_{2}^{\left(
2\right) } \\
.. & .. & .. & .. & .. & .. \\
\overrightarrow{r}_{n}^{\left( 1\right) } & t_{n}^{\left( 1\right) } &
\overrightarrow{s}_{n}^{\left( 1\right) } & \overrightarrow{r}_{n}^{\left(
2\right) } & t_{n}^{\left( 2\right) } & \overrightarrow{s}_{n}^{\left(
2\right) }
\end{array}
\]
Each row of the table lists a set of coordinated coordinates and locates a
point on the configuration space-time trajectory of the system. Taking the
two halves of the table separately, corresponding to each of the individual
particle coordinates:
\[
\begin{array}{ccc}
\overrightarrow{r}_{0}^{\left( 1\right) } & t_{0}^{\left( 1\right) } &
\overrightarrow{s}_{0}^{\left( 1\right) } \\
\overrightarrow{r}_{1}^{\left( 1\right) } & t_{1}^{\left( 1\right) } &
\overrightarrow{s}_{1}^{\left( 1\right) } \\
\overrightarrow{r}_{2}^{\left( 1\right) } & t_{2}^{\left( 1\right) } &
\overrightarrow{s}_{2}^{\left( 1\right) } \\
.. & .. & .. \\
\overrightarrow{r}_{n}^{\left( 1\right) } & t_{n}^{\left( 1\right) } &
\overrightarrow{s}_{n}^{\left( 1\right) }
\end{array}
\]
and
\[
\begin{array}{ccc}
\overrightarrow{r}_{0}^{\left( 2\right) } & t_{0}^{\left( 2\right) } &
\overrightarrow{s}_{0}^{\left( 2\right) } \\
\overrightarrow{r}_{1}^{\left( 2\right) } & t_{1}^{\left( 2\right) } &
\overrightarrow{s}_{1}^{\left( 2\right) } \\
\overrightarrow{r}_{2}^{\left( 2\right) } & t_{2}^{\left( 2\right) } &
\overrightarrow{s}_{2}^{\left( 2\right) } \\
.. & .. & .. \\
\overrightarrow{r}_{n}^{\left( 2\right) } & t_{n}^{\left( 2\right) } &
\overrightarrow{s}_{n}^{\left( 2\right) }
\end{array}
\]
yields world-lines and associated spin-vectors for each particle, but in
taking the coordinates separately one looses the information regarding the
coordination of the coordinates present in the configuration space-time
trajectory. Nonetheless, each position on the world line of each particle
has a definite spin associated with it, combinations of individual particle
coordinates from the same row of the full table will have total spin zero,
but combinations from different rows need not.

Let us consider an example. If the initial point in configuration space-time
is chosen such that $t_{0}^{\left( 1\right) }=t_{0}^{\left( 2\right) }<<t_{i}
$ , in $\Sigma ,$ then for some specific choice of $r^{\left( 1\right) }$
and $r^{\left( 2\right) },$ a configuration space-time trajectory and
particle world lines, can be calculated. For this specific choice of the
initial time coordinates the particles will reach the Stern Gerlach devices
at more or less the same time in $\Sigma $. The ``look-up'' table will then
be as shown in the following table:
\[
\begin{array}{cccccc}
\overrightarrow{r}_{0}^{\left( 1\right) } & t_{0}^{\left( 1\right) } &
\left( 0,0,0\right)  & \overrightarrow{r}_{0}^{\left( 2\right) } &
t_{0}^{\left( 2\right) } & \left( 0,0,0\right)  \\
\overrightarrow{r}_{1}^{\left( 1\right) } & t_{1}^{\left( 1\right) } &
\left( 0,0,0\right)  & \overrightarrow{r}_{1}^{\left( 2\right) } &
t_{1}^{\left( 2\right) } & \left( 0,0,0\right)  \\
\overrightarrow{r}_{i}^{\left( 1\right) } & t_{i}^{\left( 1\right) } &
\left( 0,0,+1\right)  & \overrightarrow{r}_{i}^{\left( 2\right) } &
t_{i}^{\left( 2\right) } & \left( 0,0,-1\right)  \\
.. & .. & .. & .. & .. & .. \\
\overrightarrow{r}_{n}^{\left( 1\right) } & t_{n}^{\left( 1\right) } &
\left( 0,0,+1\right)  & \overrightarrow{r}_{n}^{\left( 2\right) } &
t_{n}^{\left( 2\right) } & \left( 0,0,-1\right)
\end{array}
\]
Now if one chooses to compare values at the uncoordinated times $t^{\left(
2\right) }=t_{k}>t_{i}$ but $t^{\left( 1\right) }=t_{j}<t_{i},$ one finds
that particle two has a definite spin in the minus $z$ direction but
particle one has spin zero. However, there is no contradiction in this as no
measurement of the spin of particle one has taken place; if such a
measurement were to be carried out one would not find the value zero, but $%
\pm 1,$ and this would obviously correspond to a different
experiment in which the configuration space trajectory would have
to be calculated afresh. (This is similar to the situation in
non-relativistic de Broglie-Bohm theory wherein the values
assigned to variables are not in general the eigenvalues of the
associated operators unless a measurement process is carried out.)
In general there will be some inertial observer whose frame of
reference will correspond with $t_{k}^{\prime (2)}=t_{j}^{\prime
(1)},$ and this observer may be tempted to assume that the
particle coordinates must be coordinated on these special
equal-time hyperplanes, but this would be to adopt a
frame-dependent and hence non-relativistic rule. In our approach
the coordination of the particle coordinates has the status of a
beable, just as the positions of the particles do in the
non-relativistic de Broglie-Bohm theory. Hence for a given run of
an experiment the coordination is fixed and is no more an
arbitrary choice of an observer than the positions of the
particles are in the non-relativistic theory. For the purposes of
calculating the possible motions of the system one may choose a
variety of initial coordinated coordinates, just as one may choose
a variety of initial positions in the non-relativistic de
Broglie-Bohm theory. However, just as in the choice of initial
coordinates in the non-relativistic theory, there will be some
subset of all possible choices for $\left(
\overrightarrow{r_{0}}^{\left( 1\right) },t_{0}^{\left( 1\right)
},\overrightarrow{r}_{0}^{\left( 2\right) },t_{0}^{\left( 2\right)
}\right) $ that in fact correspond with a non-negligible
probability for a given experimental design.

Since our method of calculation is lorentz invariant, representation of the
configuration-space-time trajectory using another inertial frame of
reference involves only a simple passive transformation of the coordinated
coordinates in the look-up table, hence for a given initial position in
configuration space-time the outcome of the experiment is the same in all
inertial frames. In a given run of the experiment there will be an unknown
(and uncontrollable) configuration space-time trajectory, L, which describes
the actual evolution of the system and the manner in which the particle
coordinates are coordinated. If a different initial point in configuration
space-time, not lying on L is chosen, then the coordination of the
coordinates is different, leading to a different configuration space-time
trajectory and there should be no surprise if this leads to a different and
apparently ``contradictory'' outcome of the experiment. The theory is
deterministic in the configuration space-time in which the trajectories do
not cross. Any point lying on L, if chosen as the starting point for the
calculation will give rise to the same L and the same trajectories in
space-time.

The assumption that the coordination of the particle coordinates is to be
arbitrarily decided, in a non-lorentz-invariant way, according to the equal
time surfaces of the inertial frame chosen to describe the experiment, is
the origin of the so-called problems associated with making hidden-variable
theories of quantum mechanics lorentz invariant at the level of the hidden
variables. Hardy \cite{hardy} implicitly applies such a non-lorentz
invariant rule to demonstrate the impossibility of lorentz-invariant hidden
variables. But it should be clear that choosing a non-lorentz invariant rule
to determine the coordination of the particles' coordinates will have
non-lorentz invariant consequences.

\section{The Horne, Shimony and Zeilinger experiment}

We have chosen this experiment since a similar, double-interferometer
experiment was used by Hardy \cite{hardy} to support his argument that
hidden-variable theories of relativistic quantum theory must have
non-lorentz-invariant hidden variables. Hardy's experiment involves pair
annihilation, a feature that we would prefer to avoid, since his argument
can be discussed without this complication.

Consider the two-particle interferometer illustrated in figure 2, discussed
first by Horne, Shimony and Zeilinger \cite{HSZ} and in the context of the
non-relativistic de Broglie-Bohm theory by Dewdney and Lam \cite{dlhsz}. At $%
t=0,$ in the $\Sigma $ frame, the source $S$ emits two distinguishable non -
interacting particles, $1$ and $2$ into the spatially distinct paths $a,b,c$
and $d$. These paths are associated with the single - particle localized
wave packets $a^{(1)},b^{(2)},c^{\left( 2\right) }$and $d^{(1)}$
respectively where the bracketed superscripts 1 and 2 label the particles.
The particles are emitted so that either:

\begin{enumerate}
\item  particle 1 follows path $a$ and particle 2 follows path $c$ or,

\item  particle 1 follows path $d$ and particle 2 follows path $b$.
\end{enumerate}

Hence, in $\Sigma ,$ the wave function on the $t=0$ hypersurface can be
taken to be
\begin{equation}
\Psi \left( \overrightarrow{r}^{(1)},0;\overrightarrow{r}^{(2)},0\right) =%
\frac{1}{\sqrt{2}}\left[ a\left( \overrightarrow{r}^{\left( 1\right)
},0\right) c\left( \overrightarrow{r}^{\left( 2\right) },0\right) +d\left(
\overrightarrow{r}^{\left( 1\right) },0\right) b\left( \overrightarrow{r}%
^{\left( 2\right) },0\right) \right]  \label{eq:hszwavefunction}
\end{equation}
The particle on path $a$ experiences a variable phase shift $\phi $, whilst
the other, on path $b$, receives a variable shift $\chi $. After reflection
at the mirrors, each particle encounters a beam-splitter ( $H_{1}$ or $H_{2}$%
) through which it may be transmitted or reflected with equal probability,
regardless of the settings of the phase shifters. Each particle eventually
emerges either in the positive or the negative sense of the vertical axis,
which is designated $z.$ The outcomes for particle $1$ will be referred to
as $1^{+}$ or $1^{-}$, and similarly $2^{+}$or $2^{-}$ for particle $2$.
There are no single particle interferences, if one just looks where particle
$1$ emerges the probabilities of $1^{+}$ and $1^{-}$ are both equal to $0.5$
independently of the settings of the phase shifters. The same is true of
particle $2$. However, the joint probabilities show interference effects as
they do depend on the settings of the phase shifters as follows:
\begin{eqnarray}
P\left\{ 1^{+}2^{+}\right\} &=&\frac{1}{4}\left( 1+\cos \left( \chi -\phi
\right) \right)  \label{eq:hszprobs} \\
P\left\{ 1^{+}2^{-}\right\} &=&\frac{1}{4}\left( 1-\cos \left( \chi -\phi
\right) \right) \\
P\left\{ 1^{-}2^{+}\right\} &=&\frac{1}{4}\left( 1-\cos \left( \chi -\phi
\right) \right) \\
P\left\{ 1^{-}2-\right\} &=&\frac{1}{4}\left( 1+\cos \left( \chi -\phi
\right) \right)
\end{eqnarray}

\subsection{The configuration space-time model}

The critical aspects of the motion of the particles take place in the one
dimension $z$, perpendicular to the beam-splitters; so all features of the
evolution of the wave function, relevant to this discussion, take place in
the four-dimensional configuration space-time $(z^{\left( 1\right)
},t^{\left( 1\right) };z^{\left( 2\right) },t^{\left( 2\right) }).$ Only the
final scattering from the beam-splitters $(H_{1}$and $H_{2})$ need be
modelled in detail as the full reflections (at $M_{1}$and $M_{2})$ simply
serve to change the direction of the particles. In figure 3, we illustrate
the possible paths of the particles, in $\Sigma ,$ in a space-time $(x,z,t)$
diagram. In this frame the particles scatter from the full reflecting
mirrors at $t_{\lambda },$ reach the final beam splitters at $t_{\mu }$ and
emerge beyond the interferometers at $t_{\nu }.$ The time-dependence of the
two packets in configuration space-time, represented by the two terms in (%
\ref{eq:hszwavefunction}), are given in this model by
\begin{equation}
\Psi \left( z^{\left( 1\right) },t^{\left( 1\right) };z^{\left( 2\right)
},t^{\left( 2\right) }\right) =\frac{1}{\sqrt{2}}\left[ a\left( z^{\left(
1\right) },t^{\left( 1\right) }\right) c\left( z^{\left( 2\right)
},t^{\left( 2\right) }\right) e^{i\phi }+d\left( z^{\left( 1\right)
},t^{\left( 1\right) }\right) b\left( z^{\left( 2\right) },t^{\left(
2\right) }\right) e^{i\chi }\right]  \label{eq:hszmultitime}
\end{equation}

According to the wave function given in equation (\ref{eq:hszmultitime}) the
point $\left( z_{0}^{\left( 1\right) },t_{0}^{\left( 1\right)
};z_{0}^{\left( 2\right) },t_{0}^{\left( 2\right) }\right) $ from which we
start calculating the trajectory can be chosen arbitrarily. Let us consider
two different scenarios.

\subsubsection{Scenario 1: configuration space-time trajectories initiated
on $\left( z_{0}^{\left( 1\right) },t_{0}^{\left( 1\right) }=t_{\protect\mu
};z_{0}^{\left( 2\right) },t_{0}^{\left( 2\right) }=t_{\protect\lambda
}\right) $}

In this scenario the particles' time coordinates are such that particle 1 at
the final beam-splitter is coordinated with particle 2 at the full
reflecting mirror. $\left| \Psi \left( z_{0}^{\left( 1\right) },t_{\mu
};z_{0}^{\left( 2\right) },t_{\lambda }\right) \right| ^{2}$ is shown in
figure 4, which also illustrates, in the reduced configuration space defined
by $\left( z^{\left( 1\right) },z^{\left( 2\right) }\right) ,$ the initial
single-particle marginal distributions and the general character of the
particle trajectories. For this scenario, the main characteristics of the
particles' motions can be deduced from the fact that configuration space
trajectories do not cross. Integrating the equations of motion from specific
choices for $z^{\left( 1\right) }$ and $z^{\left( 2\right) }$ it is clear
that when particle $1$ reaches its beam splitter, the packets for particle $%
2 $ are still well separated and effectively non-overlapping. Consequently,
the two packets in configuration space are non overlapping and hence behave
independently as particle $1$ scatters from its beam splitter. In a single
instance the point representing the configuration of the system must lie in
one or other of the configuration space wave packets. If the coordinate of
particle 2 lies in $c$ then the \textit{effective} configuration space wave
function is
\begin{equation}
a\left( z^{\left( 1\right) },t_{\mu }\right) c\left( z^{\left( 2\right)
},t_{\lambda }\right) e^{i\phi }
\end{equation}
a simple product. Whereas if the coordinate of particle 2 lies in $b$ then
the effective wave function is
\begin{equation}
d\left( z^{\left( 1\right) },t_{\mu }\right) b\left( z^{\left( 2\right)
},t_{\lambda }\right) e^{i\chi }
\end{equation}
In this scenario the possible trajectories of particle 1 at its beam
splitter are just those of a single particle scattering from a beam
splitter. The non-crossing of the trajectories now applies to the individual
trajectories for particle 1 without reference to the position of particle 2
(save to locate the system in one of the quadrants of the configuration
space). If particle 1 is located in the forward part of $a(z^{\left(
1\right) },t_{\mu })$ it is transmitted and if in the trailing part it is
reflected \cite{dewdney82}. A similar argument clearly holds for the case in
which particle 2 is in $b$, but then we are concerned with the other packet
in configuration space.

After particle 1 has scattered there will be four non-overlapping packets in
configuration space. Particle 1 is leaving the interferometer whilst
particle 2 is approaching its beam splitter. The wave function can be
represented by
\begin{equation}
\Psi \left( z^{\left( 1\right) },t_{\nu };z^{\left( 2\right) },t_{\mu
}\right) =\frac{1}{2}\left[
\begin{array}{c}
\left\{ a_{r}\left( z^{\left( 1\right) },t_{\nu }\right) e^{-i\frac{\pi }{2}%
}+a_{t}\left( z^{\left( 1\right) },t_{\nu }\right) \right\} c\left(
z^{\left( 2\right) },t_{\mu }\right) e^{i\phi } \\
+\left\{ d_{r}\left( z^{\left( 1\right) },t_{\nu }\right) e^{-i\frac{\pi }{2}%
}+d_{t}\left( z^{\left( 1\right) },t_{\nu }\right) \right\} b\left(
z^{\left( 2\right) },t_{\mu }\right) e^{i\chi }
\end{array}
\right]  \label{eq:hsz2scatters}
\end{equation}
where the subscripts $r$ and $t$ represent, respectively, the reflected and
transmitted parts of the packets $a$ and $d$ . As particle 2 progresses
towards its beam splitter the four wave packets interfere at the $z^{\left(
2\right) }$ beam splitter in pairs. The functional form of $a_{r}\left(
z^{\left( 1\right) },t^{\left( 1\right) }\right) $ and $d_{t}\left(
z^{\left( 1\right) },t^{\left( 1\right) }\right) $ is identical (the
reflected part of $a$ coincides with the transmitted part of $d$) and we
represent this common function by $a_{r}d_{t}$. The same is true of $%
a_{t}\left( z^{\left( 1\right) },t^{\left( 1\right) }\right) $ and $%
d_{r}\left( z^{\left( 1\right) },t^{\left( 1\right) }\right) $ which we
represent by $a_{t}d_{r}$. Rearranging equation (\ref{eq:hsz2scatters}) we
find
\begin{equation}
\Psi \left( z^{\left( 1\right) },t_{\nu };z^{\left( 2\right) },t_{\mu
}\right) =\frac{1}{2}\left[
\begin{array}{c}
a_{r}d_{t}\left( z^{\left( 1\right) },t_{\nu }\right) \left\{ c\left(
z^{\left( 2\right) },t_{\mu }\right) e^{-i\frac{\pi }{2}}e^{i\phi }+b\left(
z^{\left( 2\right) },t_{\mu }\right) e^{i\chi }\right\} \\
+a_{t}d_{r}\left( z^{\left( 1\right) },t_{\nu }\right) \left\{ c\left(
z^{\left( 2\right) },t_{\mu }\right) e^{i\phi }+b\left( z^{\left( 2\right)
},t_{\mu }\right) e^{-i\frac{\pi }{2}}e^{i\chi }\right\}
\end{array}
\right]  \label{eq:hsz2scattersre}
\end{equation}
now the particle 1 wave packets $a_{r}d_{t}$ and $a_{t}d_{r}$ are
non-overlapping and so once more each part of the wave function behaves
independently. As is clear from equation (\ref{eq:hsz2scattersre}), the fate
of particle $2$ depends on which packet particle 1 is in and on the relative
phase of the packets $c$ and $b$. Notice that as particle 2 approaches its
beam-splitter the relative phase of the particle-2 packets $c$ and $b$
depends not just on the phase shift $\chi $ applied to particle-2 packet $b$
but also on the phase shift $\phi $ applied to particle 1 \textit{at an
arbitrarily distant location}. If, in a particular case, particle 1 is in $%
a_{r}d_{t}$ (in the positive domain of the $z^{\left( 1\right) }$ axis
corresponding with the outcome $1^{+}$) and $\chi -\phi =0,$ then all
trajectories from $c(2)$ are transmitted and all those from $b(2)$ are
reflected. These are the reduced configuration space, $\left( z^{\left(
1\right) },z^{\left( 2\right) }\right) ,$ trajectories shown in figure 4 in
the region where $z^{\left( 1\right) }$ is positive (on the right of the
diagram); they correspond with the outcome $1^{+}2^{+}$. If on the other
hand particle 1 is located in $a_{t}d_{r}$ (in the negative domain of the $%
z^{\left( 1\right) }$ axis corresponding with the outcome $1^{-}$) and $\chi
-\phi =0,$then all trajectories from $c(2)$ are reflected and all those from
$b(2)$ are transmitted. These trajectories are also shown in figure 4 in the
region where $z^{\left( 1\right) }$ is negative (on the left of the
diagram); they correspond with the outcome $1^{-}2^{-}$. Changing the phase
difference, $\chi -\phi ,$ will produce different trajectories but we need
not consider the alternatives here.

\subsubsection{Scenario 2: configuration space-time trajectories are
initiated on $\left( z_{0}^{\left( 1\right) },t_{0}^{\left( 1\right) }=t_{%
\protect\lambda };z_{0}^{\left( 2\right) },t_{0}^{\left( 2\right) }=t_{%
\protect\mu }\right) $}

In this scenario the particles' time coordinates are such that particle 2 at
the final beam-splitter is coordinated with particle 1 at the full
reflecting mirror. $\left| \Psi \left( z_{0}^{\left( 1\right) },t_{\lambda
};z_{0}^{\left( 2\right) },t_{\mu }\right) \right| ^{2}$ is plotted in the
reduced configuration space $\left( z^{\left( 1\right) },z^{\left( 2\right)
}\right) $ corresponding with $t_{0}^{\left( 1\right) }=t_{\lambda
},t_{0}^{\left( 2\right) }=t_{\mu }$ in figure 5, along with indicative
trajectories for the case in which $\chi -\phi =0$. The trajectories are
deduced in the same manner as for the alternative scenario discussed above
and the motion of particle 2 at its beam splitter is independent of the
location of particle 1 within the appropriate configuration space packet.
Particles in the forward parts of the packets $b\left( 2\right) $ and $%
c\left( 2\right) $ are transmitted whilst those in the trailing parts are
reflected. The fate of particle 1 at its beam splitter then depends on the
position of particle 2 after passage through the beam splitter.

\subsubsection{Comparison of the scenarios}

Consider the case in which particle 1 is located in the front part of packet
$a(1)$, in scenario 1 this particle will definitely be transmitted
(irrespective of the position of particle 2 in its packet), whereas in
scenario 2, if particle 2 is in the rear part of $c\left( 2\right) ,$ then
particle 1 will be reflected. If in a given single run of the experiment the
choice between the different scenarios is an arbitrary choice of the
``observer'' (rather than a ``beable'' determined by the hidden-variable
theory) then there is a contradiction: this is what happens in the
multi-simultaneity theory and forms the basis of Hardy's argument. Hardy in
fact gives no explicit rule for coordinating the particle time coordinates,
but implicitly relies on the rule that the coordinates are to be coordinated
along the equal-time hypersurfaces of the arbitrary frame of reference
chosen to describe the experiment. This is similar to multi-simultaneity
theory which also relies on a non-Lorentz invariant rule.

According to multi-simultaneity theory, the frame of reference in which the
two particles are coordinated along the equal time hypersurfaces is chosen
to be the rest frame of the massive apparatuses with which the particles
interact. The apparatus components in the two distant wings of the HSZ
experiment can be put in such a state of motion that, according to
multisimultaneity, both of the scenarios that we have discussed apply at
once. It is then conjectured that this clash is reflected in nature by the
loss of correlations.

Bohm's preferred-frame theory rejects the joint applicability of the two
scenarios and argues that there is only one preferred frame, $\Sigma
^{\prime }$, in which the results can be calculated correctly. The preferred
frame of reference may, or may not, coincide with a frame which has $%
t_{\lambda }^{\prime (1) }=t_{\mu }^{\prime (2)}$, but at most
only one of the scenarios we can calculate can describe the
correct motion. The preferred frame is undetectable
experimentally.

In our approach the different possible motions for the particles in the two
scenarios arise, not as a result of different arbitrary choices of
experimenters, but from different initial conditions, $\left( z_{0}^{\left(
1\right) },t_{0}^{\left( 1\right) };z_{0}^{\left( 2\right) },t_{0}^{\left(
2\right) }\right) ,$ that may, or may not, have been physically realized for
a given experimental situation and, given the initial values (which
determines the coordination of the trajectories) our calculation yields the
same motion regardless of our choice of frame of reference. The two
scenarios correspond to different and exclusive ``experimental runs''; they
are not equally valid descriptions of one and the same experimental run. If,
for example, the wave packets in fact emerge simultaneously in a given frame
of reference, then this determines the ensemble of possible, physically
realizable, initial configuration space-time points (and hence
coordinations) for this particular experiment. Observers in frames of
reference in relative motion will judge simultaneity differently, but this
has nothing to do with the actual set of initial points in configuration
space-time that are consistent with the experimental conditions or the way
in which the points on the individual particle trajectories are in fact
coordinated according to our lorentz invariant rule. So, a given instance of
the experiment will have a corresponding lorentz-invariant trajectory in the
configuration space-time, but, of course, we can neither know in advance
which trajectory is actualized nor control the hidden variables. Detection
of the particles at the space-time points $\left( z^{\left( 1\right)
},t^{\left( 1\right) }),(z^{\left( 2\right) },t^{\left( 2\right) }\right) $
does not reveal sufficient information to determine the configuration-space
trajectory which the particles in fact followed; additionally one needs to
know whether the particle trajectories were coordinated at these points. In
other words, the particles may have been detected at $\left( z^{\left(
1\right) },t^{\left( 1\right) })(z^{\left( 2\right) },t^{\left( 2\right)
}\right) $ but this does not reveal whether these points are coordinated.
Initiating a configuration-space-time trajectory from the point $\left(
z^{\left( 1\right) },t^{\left( 1\right) };z^{\left( 2\right) },t^{\left(
2\right) }\right) $, with the idea of retrodicting the system trajectory,
would assume that the individual trajectories are coordinated by this point,
which need not be the case.

\section{Conclusion}

The apparent lack of a covariant extension of the de Broglie-Bohm
hidden-variable theory of relativistic quantum mechanics, which respects
lorentz invariance at the level of the hidden variables, has often been
cited as a major problem. In this paper we have shown how our
lorentz-invariant extension of the de Broglie-Bohm approach can be applied
to EPR - type experiments to give a unique and unambiguous lorentz invariant
description of the individual processes which underlie the statistical
predictions of quantum theory. Our approach is in accord with
non-relativistic de Broglie-Bohm theory in which, given a system's wave
function, one only needs to specify an initial position in the system's
configuration space-time in order to completely determine the motion.
However, in the relativistic theory the specification of the initial
position is no longer sufficient, one also needs to specify a rule according
to which the particles' coordinates are coordinated (beyond the initial
point) in integrating the equations of motion. The initial point, plus the
rule, leads to a unique trajectory in the system's configuration space-time
for each experimental run. Hence for any experiment one has a pair of unique
world lines along which the coordinates are coordinated according to the
given configuration space-time trajectory. One can then choose to describe
the well-defined configuration space-time trajectory using any inertial
frame that one likes, this choice is arbitrary just as one would expect in a
relativistic theory.

The adoption of non-lorentz invariant (or inertial frame dependent) rules of
coordination leads to contradictions, as Hardy has demonstrated and as is
clear in multisimultaneity theory. In this paper we have shown how the
ambiguities discussed by Hardy (and others) can be avoided in our approach
by the use of a natural lorentz invariant theory and how the fundamental
notion of lorentz invariant ``beables'' can be retained.

\section{Figure Captions}

\begin{description}
\item  Figure 1. Space -time paths in the EPR-Bohm experiment. The
measurements are carried out simultaneously in this frame at
$t=t_i$. (Arbitrary units).

\item  Figure 2. The Horne-Shimony-Zeilinger (HSZ) experiment.
Particle 1 encounters a beam-splitter ($H_{1}$) at $x=-10$, whilst
particle two encounters a beam splitter ($H_{2}$) at $x=10$
(arbitrary units).

\item  Figure 3. Space-time paths in the HSZ experiment in the
$\Sigma$ frame. "ardt" labels the exit beam composed of the
reflected part of packet $a(1)$ and the transmitted part of packet
$d(1)$. The other labels have similar interpretation.

\item  Figure 4. $\left| \Psi \left( z^{\left( 1\right) },t_{\mu };z^{\left(
2\right) },t_{\lambda }\right) \right| ^{2}$in the reduced configuration
space $\left( z^{\left( 1\right) },z^{\left( 2\right) }\right) $ shown with
indicative trajectories for the case in which $\chi -\phi =0$ for scenario 1
in which the trajectories are initiated in positions for which $t^{\left(
1\right) }=t_{\mu }$ and $t^{\left( 2\right) }=t_{\lambda }.$

\item  Figure 5. $\left| \Psi \left( z^{\left( 1\right) },t_{\lambda
};z^{\left( 2\right) },t_{\mu }\right) \right| ^{2}$in the reduced
configuration space $\left( z^{\left( 1\right) },z^{\left( 2\right) }\right)
$ shown with indicative trajectories for the case in which $\chi -\phi =0$%
for scenario 2 in which the trajectories are initiated in positions for
which $t^{\left( 1\right) }=t_{\lambda }$ and $t^{\left( 2\right) }=t_{\mu
}. $
\end{description}

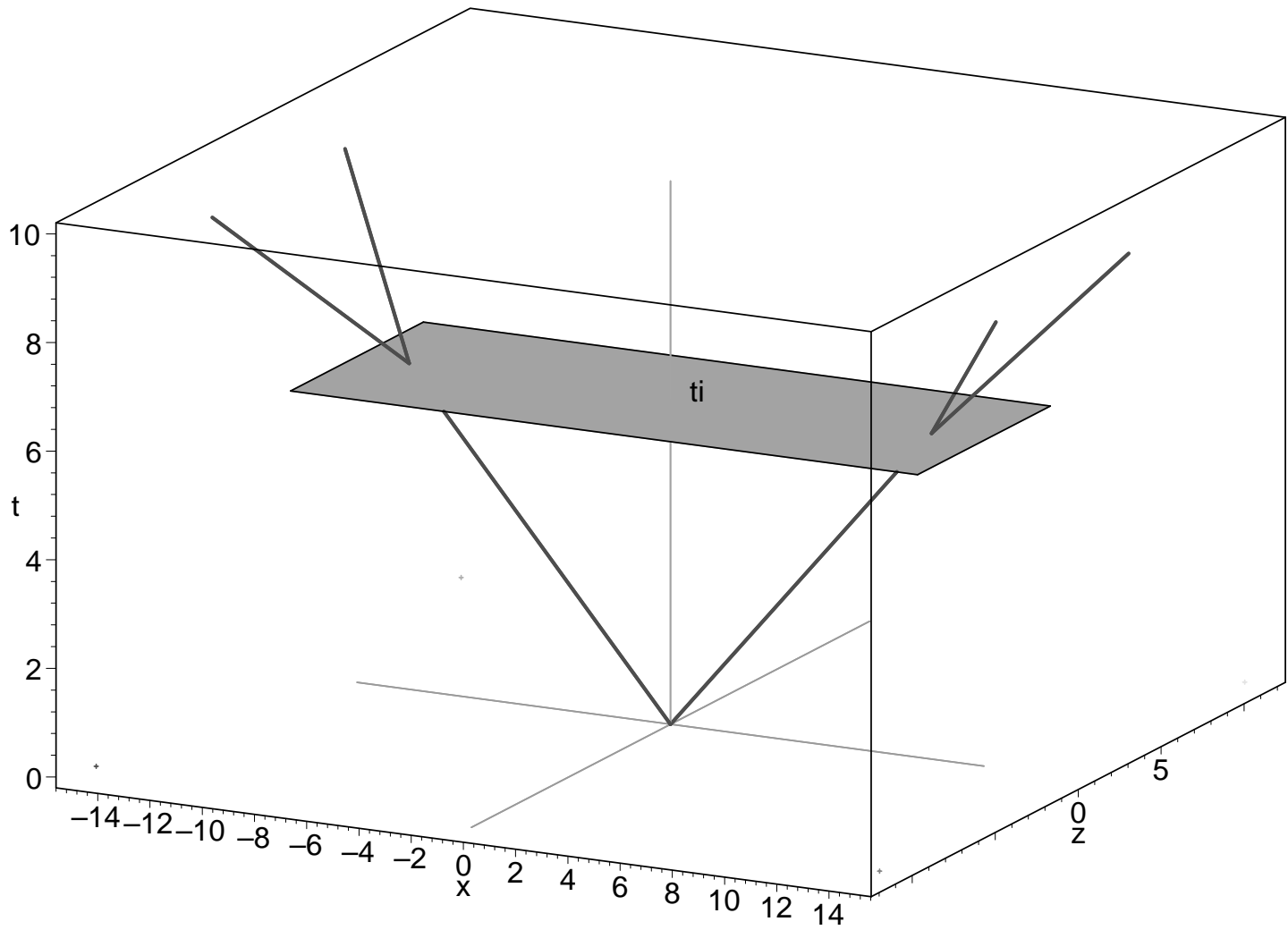
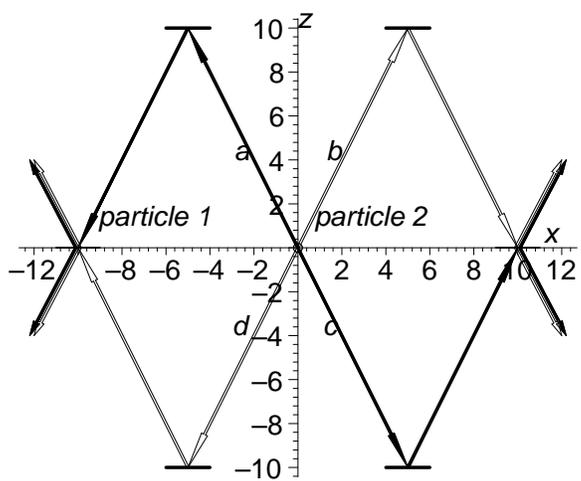
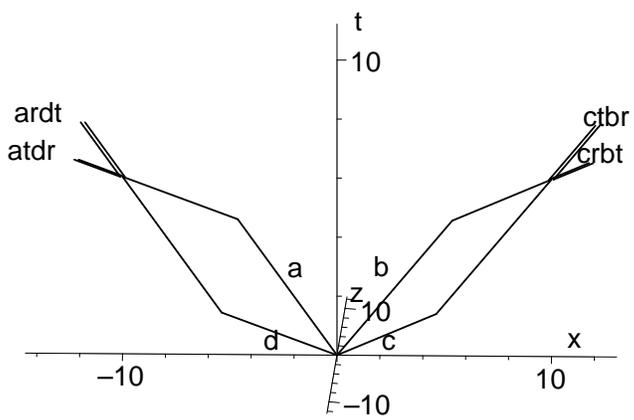
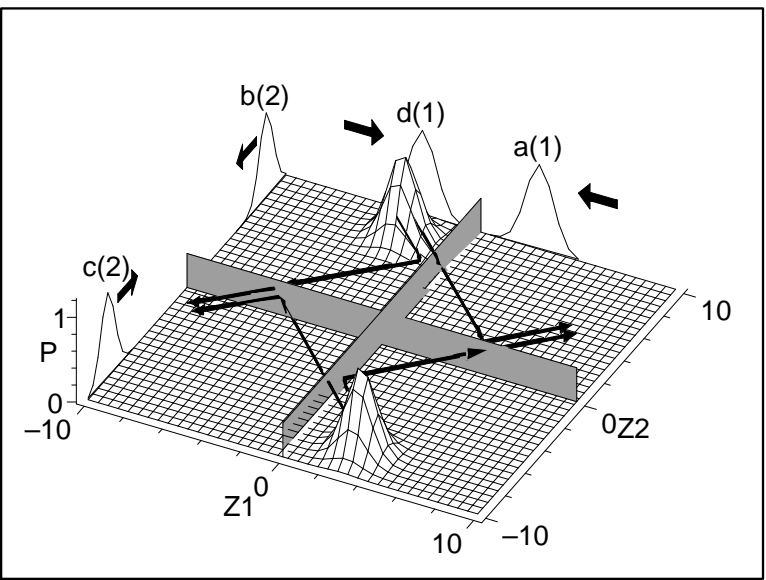
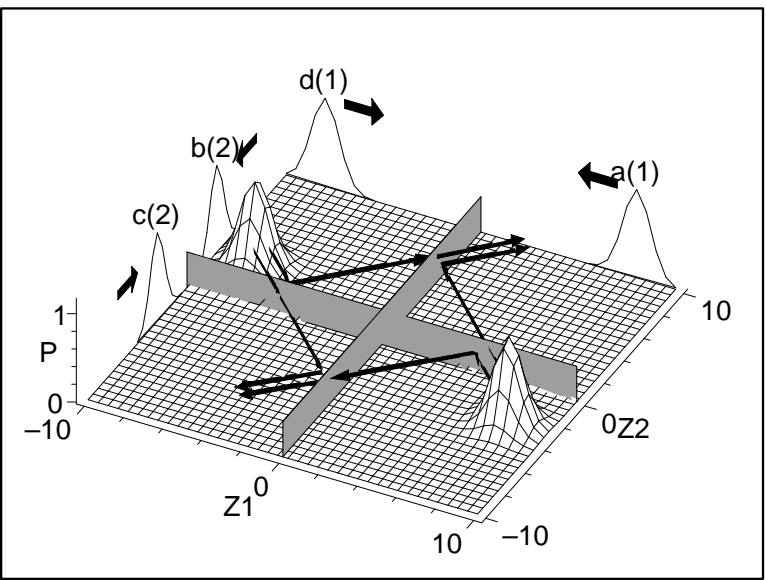
\end{document}